\documentclass[12pt]{iopart}
\usepackage{latexsym}
\usepackage{graphics}
\usepackage{graphicx,psfrag}
\begin{document}
\title{ Creation and annihilation of fluxons in ac-driven semi-annular Josephson junction  }
\author{Chitra R N$^\dag$ and V C Kuriakose$^\ddag$}

\address{Department of Physics, Cochin University of Science
and
Technology, Kochi, 682022} %
\ead{$^\dag$rchitra.r@gmail.com, $^\ddag$vck@cusat.ac.in}

\begin{abstract}The dynamical behavior of a fluxon in a
semi-annular long Josephson junction in the presence of an
ac-drive is studied. The non-uniformity due to the non uniform
distribution of bias current is investigated. The oscillating
potential is found to increase the depinning current. Finite
difference method is used for numerical analysis and the
response of the system to the ac-bias is studied. The creation
and annihilation of fluxon is also demonstrated numerically for
the first, second and third Zero-Field step cases.
\end{abstract}
 \maketitle
\section{Introduction}
A fluxon in long Josephson junctions (LJJ) is a well-known
physical example of a sine-Gordon fluxon. Fluxons, endemic to
LJJs, have been employed in the fabrication of devices like
constant voltage standards \cite{kau:77,ngu:07}, flux flow
oscillators \cite{nag:83,jaw:08}, logic gates
\cite{nak:99,sha:99}and also in qubits \cite{tan:01,mak:01}.
LJJs of various geometries have been thoroughly studied both
experimentally and theoretically in the past. Fluxon dynamical
properties like fluxon pinning \cite{ver:97}, fluxon trapping
\cite{kei:96}, and phase locked states have been studied for
rectangular \cite{mcl:78,lom:82} and annular \cite{ust:92} LJJs.

The non rectangular Josephson junction has been in the focus of
fluxon dynamics study in recent years because of the non
uniformity caused by the shape. Semicircular geometry for
Josephson junction has been proposed and fluxon dynamics has
been studied both analytically and numerically and its various
applications has been discussed \cite{sha:02}. It has been shown
that in the presence of an external magnetic field applied
parallel to the dielectric barrier of such a geometry, the ends
of the junction has opposite polarities and because of that
opposite polarity fluxons can enter the junction from the ends
under a properly biased dc current. If the direction of the
current is reversed, flux penetration and progression is not
possible and flux free state exists in the junction. This unique
phenomenon cannot be achieved in any other geometry and thus
this junction behaves as a perfect diode. The effect of in-plane
static and rf-magnetic field on fluxon dynamics in a semiannular
Josephson junction has also been studied \cite{sha:04}. The
response of a fluxon to an ac-drive has investigated by several
authors. It was shown that in a system with periodic boundary
condition average progrssive motion of fluxon commenses after
the amplitude of the ac drive exceeds a certain threshold value
\cite{gol:02}. Complex switching distributions has been obtained
for ac-driven annular JJs and theoretical explanation has been
provided for the multipeaked experimental observations
\cite{gron:04}. The behavior of fluxon under two ac forces has
been studied and it was shown that the direction of motion of
fluxon is dependent on ratio of frequencies, amplitudes and
phases of the harmonic forces \cite{mor:06}. In this work, we
study the effect of an ac-bias applied in the plane of a
semi-annular Josephson junction. in section II we discuss the
equation representing the junction and arrive at an expression
for the potential of the junction. The numerical results are
also presented. In section III we demonstrate creation and
annihilation of fluxons in semiannular JJs in the presence of an
ac bias and an external magnetic field. Section IV deals with
the results and discussion.
\section{Perturbation analysis of a fluxon in a semiannular junction}
The dynamical equation for a semiannular LJJ in a harmonically
oscillating field applied in its plane is
\begin{equation}\label{singannu}
\varphi_{tt}-\varphi_{xx}+\sin\varphi=
-\alpha\varphi_t+\beta\varphi_{xxt}-
b\cos(kx)-\gamma+i_0\sin(\omega t)
\end{equation}
where $\varphi(x,t)$ is the superconducting phase difference
between the electrodes of the junction with the spatial
coordinate $x$ normalized to $ \lambda_J$, the Josephson
penetration depth and time $t$ normalized to the inverse plasma
frequency $\omega_0^{-1}$and
$\omega_0=\frac{\widetilde{c}}{\lambda_J}$, $\widetilde{c}$
being the maximum velocity of the electromagnetic waves in the
junction. $R$ is resistance per unit length, $L_p$ is the
inductance per unit length, $C$ is the capacitance per unit
length, and $\gamma=\frac{j}{j_0}$is the normalized amplitude of
a dc bias normalized to maximum Josephson current $j_0$ and $i_0
\sin(\omega t)$ is the applied ac biasing.  $\alpha$ is the
quasiparticle tunneling loss and  $\beta$ is the surface loss
term in the electrodes and their values vary from $0.001$ to
$0.3$ in experiments. The surface loss term is important and it
will dominate fluxon propagation in some cases. The term $b
sin(kx)$ is due to the semicircular geometry of the junction and
$k=\frac{\pi}{l}$ and $b=2\pi\lambda_J\Delta
Bk/\Phi_0=2k(B/B_{c1})$, where
$B_{c1}=\frac{\Phi_0}{\pi\Delta\lambda_J}$ is the first critical
field of the Josephson junction. $\Phi_0=\frac{h}{2e}$ is the
flux quantum and its value is $2.064\times10^{-15}$. The extra
term $b sin(kx)$ corresponds to a force that drives fluxons
towards the left and anti fluxon towards the right. Thus in the
absence of an external field a flux free state will exist in the
junction as any static trapped fluxon present in the junction
will be removed \cite{sha:04}. In the absence of any
perturbation (\ref{singannu}) reduces to simple sine-Gordon
equation with fluxon solution given by
\begin{equation}\label{sol}
\varphi(x,t)=4 \tan^{-1}[\exp\frac{\sigma(x-X)}{\sqrt{1-u^{2}}}]
\end{equation} where $u$ is the velocity of the fluxon and $X=ut+x_0$ is the
instantaneous location of the fluxon. $\sigma=\pm1$ is the
polarity of the flux quantum (which means there are two
orientations for the fluxon). A quantum of flux in one direction
is called the kink (fluxon) fluxon and that in other direction
is called antikink fluxon (antifluxon).
\subsection{Expression for potential function}
The Lagrangian density of Eq. \ref{singannu} with $\gamma=\alpha=i_0=\beta=0$ is
\begin{equation}
 \textbf{L}=\frac{1}{2}\varphi_t^2-\frac{1}{2}\left(\varphi_x-\frac{b}{k}\sin(kx)\right)^2-1+\cos\varphi
\end{equation}
where the first term is the kinetic energy associated with the
energy density of the electric field, the second term accounts
for the potential energy density associated with the magnetic
field and the third term represents the Josephson coupling
energy density. From the potential energy density term, the
change in potential energy due to the combined effect of fluxon
motion  and the applied field can be determined by integrating
the term $-\frac{b}{k}\sin(kx)\varphi_x$ over the length of the
junction \cite{sha:04}. The fluxon induced potential as a
function of the fluxon coordinate X may be calculated as
\begin{equation}
 U(X)=-\frac{b}{k}\int_{-\infty}^{\infty}\sin(kx)\varphi_xdx
\label{potenunbiased}
\end{equation}
The integration over $-\infty$ to $\infty$ may be justified as the length of the junction is very large as compared to the size of the fluxon. Substituting Eq. \ref{sol} in Eq. \ref{potenunbiased} and integrating we get the expression for potential as
\begin{equation}
 \label{potenunbiafinal}
U(X)=-2 b l \sec h
\left(\frac{\pi^2}{2l}\sqrt{1-u^2}\right)\sin(kX)
\end{equation}
For $u\simeq0$ we can write
\begin{equation}
 \label{potenunbiafinal1}
U(X)=-2 b l \sec h \left(\frac{\pi^2}{2l}\right)\sin(kX)
\end{equation}
which has a potential well form with the depth of the well
depending on $b$ and $l.$ The vortex will be pinned to the
potential minima as long as the bias current is smaller than the
depinning current. The pinned state of a vortex corresponds to a
zero voltage state.

Now we  arrive at an expression for the potential function of
the perturbed system. The Hamiltonian of the system can be
written as a combination of the Hamiltonian of the unperturbed
sine Gordon part plus the hamiltonian of the perturbation
part~\cite{mcl:78}. Energy of the unperturbed sine-Gordon system
is
\begin{equation}\label{ham}
H^{SG}=\int_{-\infty}^{\infty}[\frac{1}{2}(\varphi_t^{2}+\varphi_x^{2}+1-cos\varphi)]dx
\end{equation}
Substituting(\ref{sol}) in (\ref{ham}) we get
\begin{equation}\label{rham}
\frac{d}{dt}H^{SG}=8u(1-u^2)^{-3/2}\frac{du}{dt}
\end{equation}
Due to the perturbational part, energy is dissipated and rate of
dissipation is given as
\begin{eqnarray}\label{dis}
\frac{d}{dt}(H^{p})&=&[\varphi_x\varphi_t]_{-\infty}^{\infty}-\\\nonumber&\int_{-\infty}^{\infty}&({\alpha\varphi_t^2+\beta\varphi_{x
t}^2+[b \cos(kx)+\gamma+i_0\sin(\omega t)]\varphi_t})dx
\end{eqnarray}
Here the first term on the right hand side accounts for the
boundary conditions and vanishes. Substituting (\ref{sol}) in
above equation we obtain the equation for rate of dissipation as

\begin{eqnarray}\label{dham}
\frac{d}{dt}(H^{p})&=& 2 \pi u \left(\gamma+i_0\sin(\omega
t)\right)-\frac{8 \alpha u^2}{\sqrt{1-u^2}}\\\nonumber&-&\frac{8
\beta u^2}{3 (1-u^2)^{3/2}}-2 \pi b u \sec
h(\frac{\pi^2\sqrt{1-u^2}}{2l})\cos(kX)
\end{eqnarray}
Following the perturbation analysis we get
\begin{eqnarray}\label{velrate}
\frac{du}{dt}&=& \frac{\pi}{4} \left(\gamma+i_0\sin(\omega
t)\right)\left(1-u^2\right)^{3/2}- \alpha u
\left(1-u^2\right)\\\nonumber&-&\frac{1}{3 }\beta u
-\frac{\pi}{4} b \left(1-u^2\right)^{3/2} \sec
h(\frac{\pi^2\sqrt{1-u^2}}{2l})\cos(kX)
\end{eqnarray}
Eq. \ref{velrate} describes the effect of perturbations on the
vortex velocity. The first term represents the effect of applied
biasing, the second and the third term represents dissipation
and fourth term is the effect of the external magnetic field on
the semicircular geometry.

In perturbational analysis, a vortex is considered as a
non-relativistic particle of rest mass $m_0=8$ moving in one
dimension. Therefore the effective potential can be obtained
using the force relation
\begin{equation}\label{potential}
\frac{\partial U_{eff}}{\partial X}=-m_0 \frac{du}{dt}
\end{equation}
Substituting Eq. \ref{velrate} in Eq. \ref{potential} and integration yields to the expression for effective potential of the form
\begin{equation}\label{potential1}
 U_{eff} (X_0)=-2 b l \sec h (\frac{\pi^2}{2 l}) \sin(k X_0)-2 \pi (\gamma+i_0 \sin(\omega t)) X_0
\end{equation}

\begin{figure}[tbh]
\centering
\includegraphics[width=6cm]{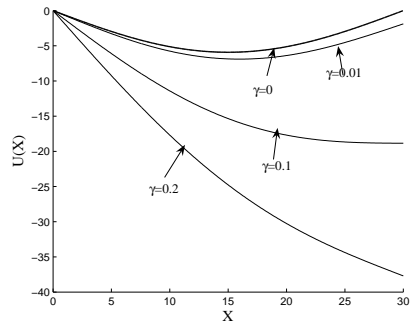}
\caption{Potential well form for a JJ of length l=15.
Other parameter values are $b=0.1$, $i_0=0.$ The $\gamma$ value is increasing from top to bottom line.} %
\label{potenwell}
\end{figure}

\begin{figure}[tbh]
\centering
\includegraphics[width=6cm]{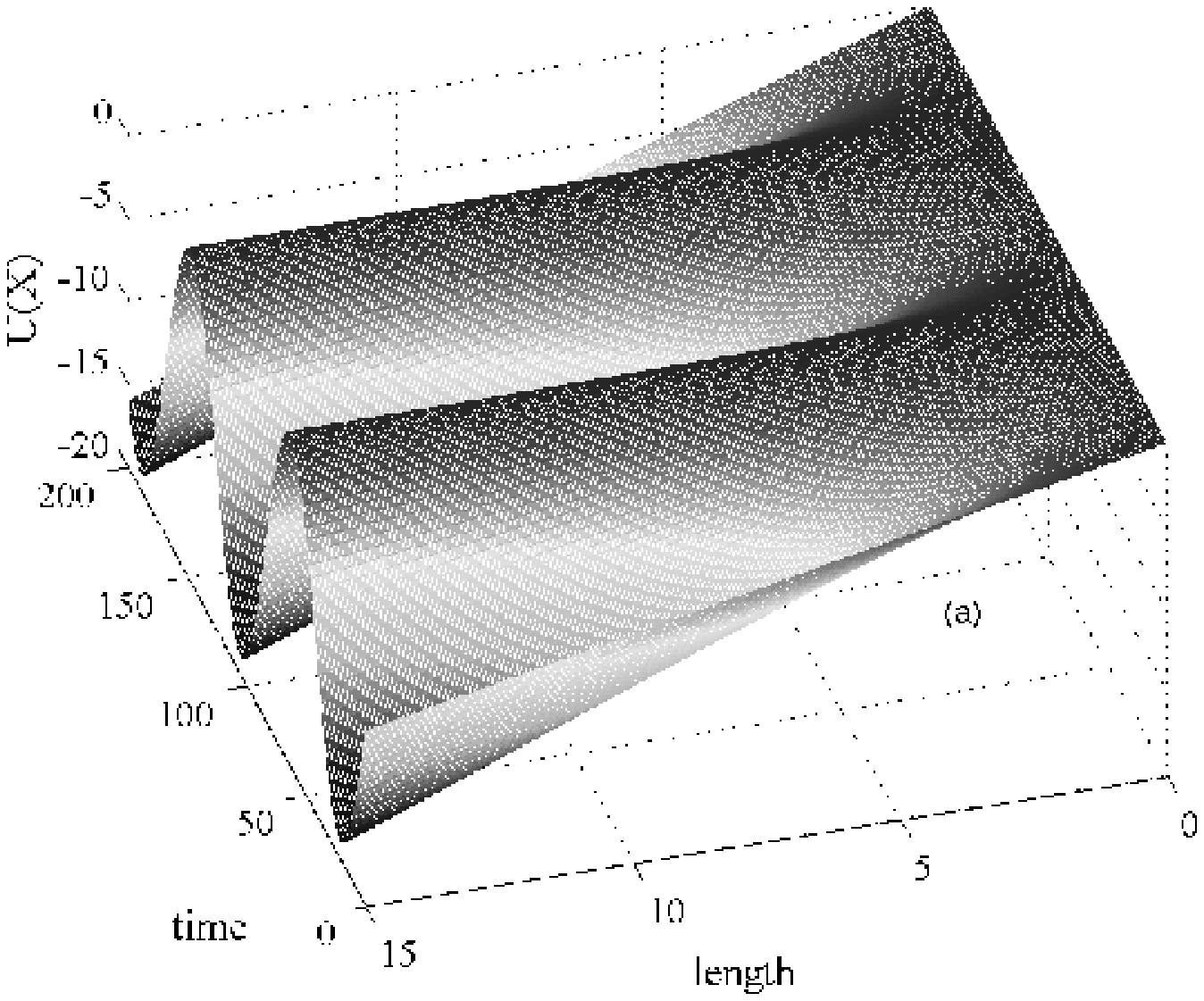}
\hspace{1cm}
\includegraphics[width=6cm]{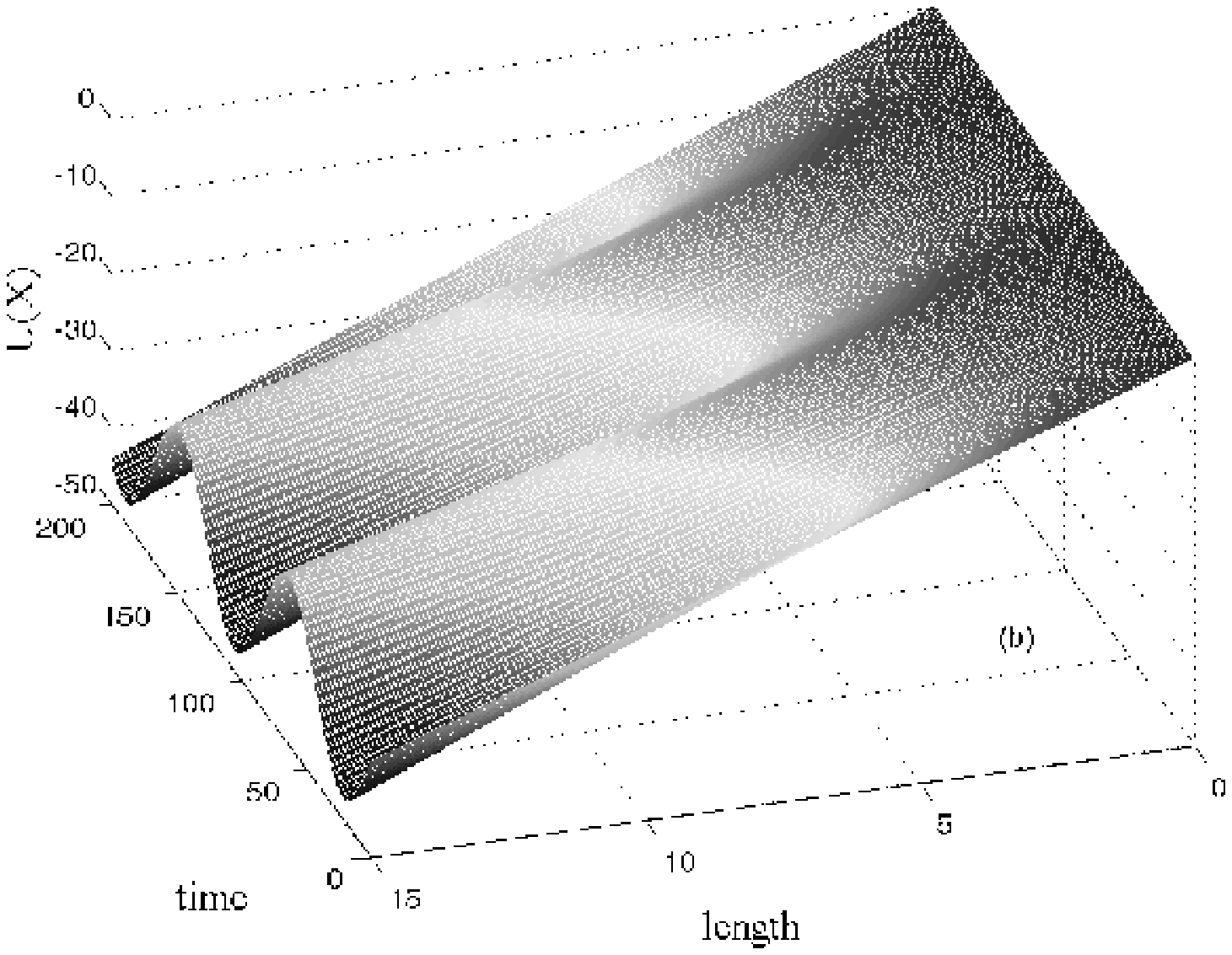}
\caption{Form of the oscillating potential for a JJ of length
l=15. Parameter values are $b=0.1$, $i_0=0.2,\omega=0.3,
\gamma=0.1$. a) applied dc bias is $\gamma=0.1$
b) $\gamma=0.4$} %
\label{potentime}
\end{figure}

The potential energy function $ U_{eff} (X)$  has a well form in
the absence of external biasing and the fluxon will remain
pinned to the centre of the junction under such a potential. As
the biasing is increased the potential gets tilted finally
favoring the motion of the vortex. The dc bias at which the zero
voltage switches to a finite voltage is called the depinning
current. Fig. \ref{potenwell} shows the form of the potential
for $b=0.1$ for a junction of length $l=15$ for different
external biasing. It can be seen that in the presence of an
external bias the potential gets tilted favoring motion of the
fluxon. While moving through such a potential, the
fluxon(antifluxon) after bouncing from the edge turns into an
antifluxon (fluxon) and hence will move in the opposite
direction.
In the presence of an ac, the potential gets oscillating with a
frequency equal to the frequency of the applied field and the
shape of the potential depends on the amplitude of the applied
ac and dc biasing. In the presence of external ac biasing along
with the dc, the potential gets time varying as shown in
fig\ref{potentime}.
In Fig. \ref{potentime}(a), the applied ac has an amplitude of
$0.1$ and it can be seen that as time goes on, the potential
gets a well form in a time period equal to the period of
oscillation of the applied field. The applied dc bias is $0.1$
in this case. Hence progressive motion of fluxon does not occur
for this value of biasing as the average velocity of a fluxon
moving in such a potential would be zero. However as the
dc-biasing is increased though the form of the potential is
still oscillating, there is a definite tilt which makes
progressive motion of fluxon possible. The response of the
system to such a potential can be investigated by measuring the
velocity of the fluxon in the potential. Thus when an ac biasing
is applied in such a form to the semi-annular JJs, progressive
motion occurs only when dc bias vale  exceeds certain threshold
value.
\subsection{Numerical Results}
To solve Eq. \ref{singannu} numerically,  we use an explicit
method treating $\phi_{xx}$ with a five point, $\phi_{tt}$ with
a three point and $\phi_{t}$ with a two point finite difference
method. The boundary conditions are treated by the introduction
of imaginary points and the corresponding finite difference
equation is solved using standard tridiagonal
algorithm\cite{paul}. Numerical simulations are carried out on
the JJ of normalized length (l=15). The time step was taken as
$0.0125$ and the space step was $0.025.$ The numerical results
were checked by systematically halving and doubling the time
steps and space steps. Details of the simulation can be obtained
from \cite{lom:82,sha:02}. After the simulation of the phase
dynamics for a transient time, we calculate the average voltage
V for a time interval T to be
$$V=\frac{1}{T}\int_0^T\varphi_t dt=\frac{\varphi(T)-\varphi(0)}{T}$$
Also for the faster convergence of the averaging procedure, the
phases  $\varphi(x)$ in the equation were averaged over the
length of the junction. The spatial averaging increases the
accuracy in the calculation of the voltages in cases where the
 the time period over which integration is made is not an exact multiple
of the time period of oscillation. Once the voltage averaging
for a current $\gamma$ is complete, it is increased in small
steps of $0.01$ to calculate the next point of the
characteristic graph. The average velocity of the fluxons can be
calculated from the average voltage using the relation $u=V(l/2
\pi)$

\begin{figure}[tbh]
\centering
\includegraphics[width=8cm]{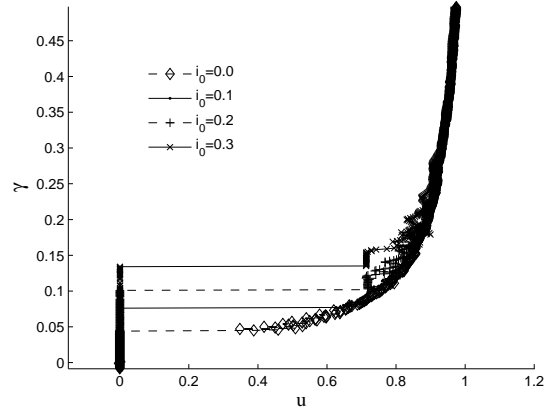}
\caption{The velocity- bias characteristics of a LJJ of length
l=15 with no external magnetic field applied. Other parameter values are $\omega=0.3,\beta=0.02,\alpha=0.05$} %
\label{nomagnet}
\end{figure}

Taking $\beta$ to be zero, the velocity change with increase in
dc biasing is observed. Fig.\ref{nomagnet} shows the velocity
change with dc biasing for different values of amplitude of the
ac biasing.  In the presence of ac biasing the averaging
interval T was taken as a multiple of the ac drive's period $2
\pi /\omega$ \cite{gol:02}. If an ac-biasing is present, the
depinning current is found to increase which can be seen from
Fig.\ref{nomagnet}. Constant voltage steps are observed for an
ac-bias of amplitude $0.2$ and $0.3.$

\begin{figure}[tbh]
\centering
\includegraphics[width=8cm]{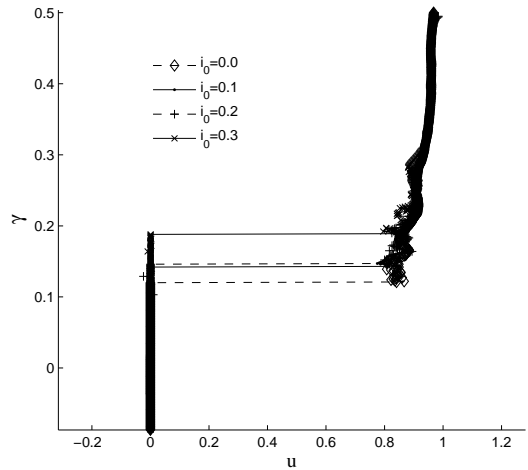}
\caption{The velocity- bias characteristics of a LJJ of length
l=15 in the presence of an external magnetic field $b=0.1$.
} %
\label{magnetb1}
\end{figure}

In the presence of external magnetic fields, the velocity versus
dc bias is shown in Fig. \ref{magnetb1}.  The value of dc bias
to cause a finite velocity for the fluxon in a JJ with a
magnetic field of $b=0.1$ and no ac-biasing is $0.1$  while for
$b=0$ the depinning current is $0.04$. Thus the external
magnetic filed also increases the depinning current value.
\begin{figure}[tbh]
\centering
\includegraphics[width=8cm]{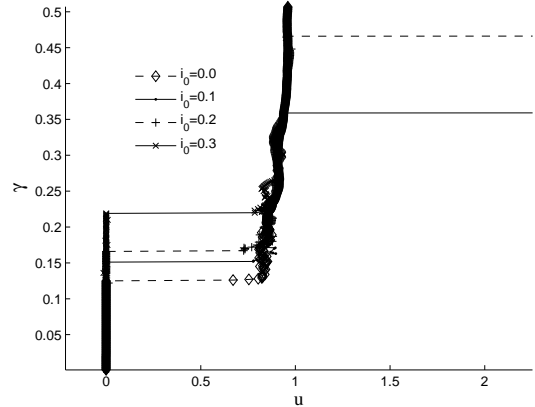}
\caption{The velocity- bias characteristics of a LJJ of length
l=15 in the presence of an external magnetic field $b=0.1$. The
damping parameter $\beta=0.035$
} %
\label{b1dam}
\end{figure}
The depinning current increases for higher values of damping
parameter $\beta=0.035.$ The depinning current to be applied to
the semi-annular JJ in the absence of ac, and a magnetic field
of $b=0.1$ is $0.125$. Also for an ac bias of amplitude $0.2$
and $0.3$ the velocity shoots to a  higher value even for a
dc-braising of $0.46$ and $0.36$. For dc bias values of more
than these values, quasiperiodic or  chaotic motion may exist in
the system.

\section{Creation and Annihilation of Fluxons}
An annular LJJ preserves the number of trapped fluxons in it.
However in an open ended geometry the number of fluxons is not a
conserved quantity. In this section we investigate the creation
and annihilation of fluxons in semiannular JJ with open boundary
conditions in the presence of an external field and an ac and dc
biasing.
\subsection{First Fiske Step}
The collision of fluxons with localized obstacles leads to
creation and annihilation of fluxons. The fluxon creation and
annihilation process for a single kink solution as input is
described here. A kink solution is launched from the centre  of
the junction with an initial velocity of $v=0.6$. For each value
of biasing the fluxon is allowed to propagate for some time in
order to stabilize its motion in the junction. The kink fluxon
gets reflected from the boundaries and moves on till
$\gamma=0.57$. Above this biasing, no solitonic propagation is
observed for an external magnetic field of strength $0.1$.

However, the presence of an ac bias creation and annihilation of
fluxon was observed for values of dc which gave one fluxon
solution earlier.
\begin{figure}[tbh]
\centering
\includegraphics[width=6cm]{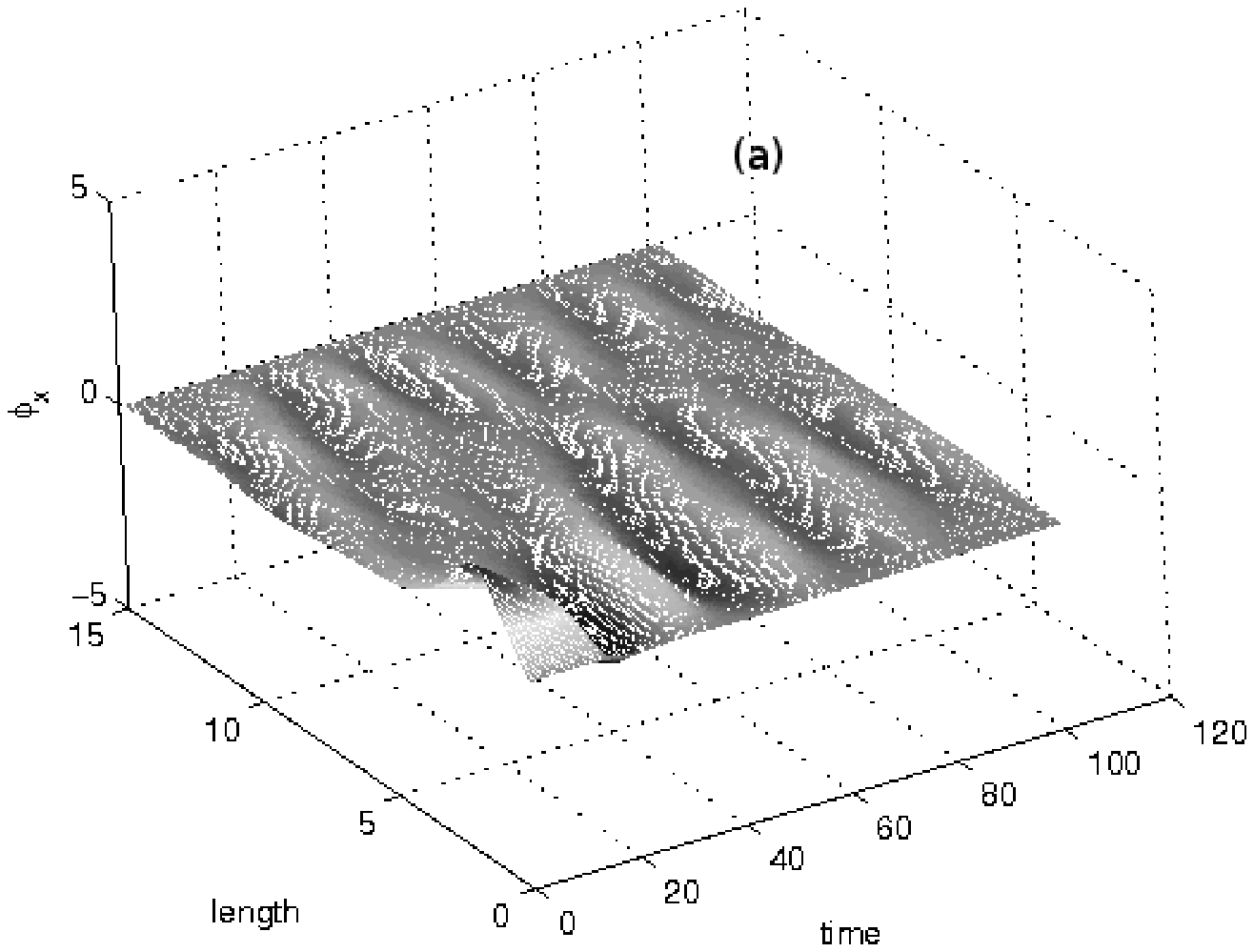}
\hspace{1cm}
\includegraphics[width=6cm]{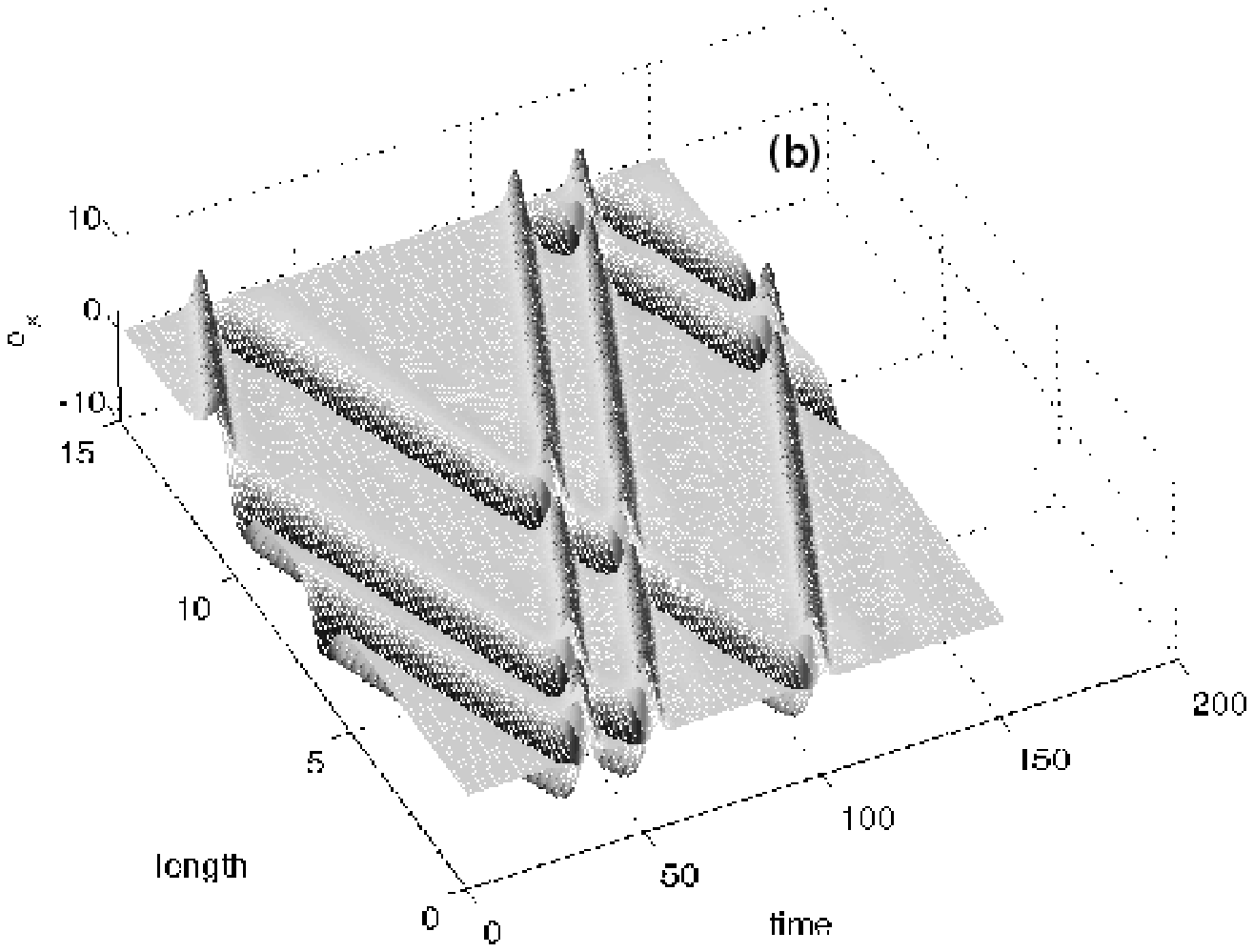}
\caption{(a)The pattern shows annhilation of fluxon propagating
in a JJ with l=15 for a dc bias of $\gamma =0.1$ and
$i_{ac}=0.2$. (b) Creation of fluxon with $\gamma =0.1$ and
$i_{ac}=0.1$
Other parameter values are $ \omega=0.3,\beta=0.02,\alpha=0.05, b=0.1$} %
\label{annihilationf}
\end{figure}
For a $\gamma$ value of $0.1$ the fluxon propagates through the
semiannular junction, while an ac bias of $0.2$ destroys the
fluxon as can be seen from Fig. \ref{annihilationf}(a).
Similarly Fig. \ref{annihilationf}(b) shows that a fluxon is
created for $\gamma =0.5$ and $i_{ac}=0.1$.
\subsection{Second Fiske Step}
\begin{figure}[tbh]
\centering
\includegraphics[width=6cm]{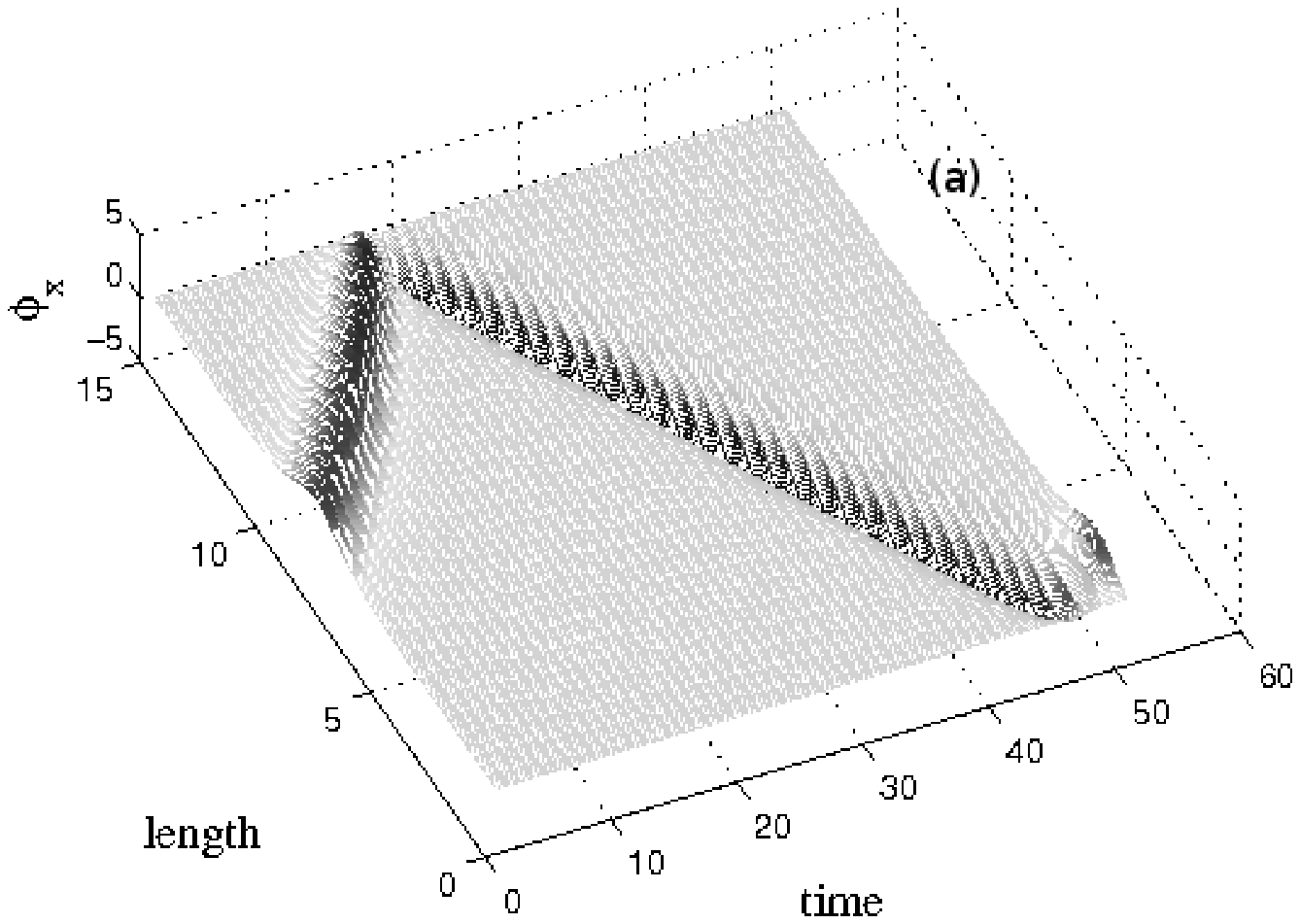}
\hspace{1cm}
\includegraphics[width=6cm]{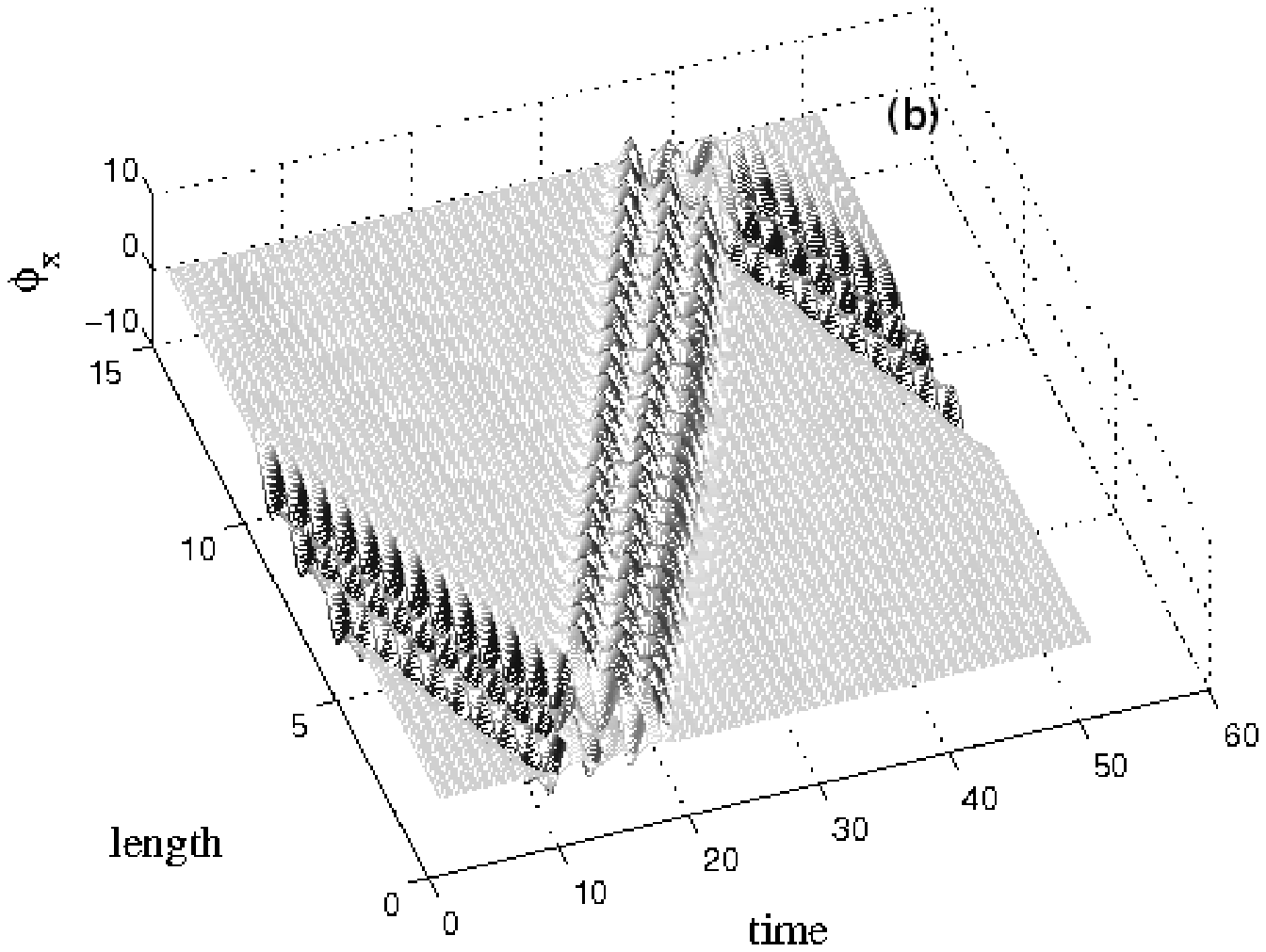}
\caption{The pattern shows single fluxon propagating in a JJ
with l=15 for  $idc=0.1$ for 2 fluxon input.(b)Creation of a fourth fluxon. $iac=0.1$, $idc=0.41$} %
\label{annihilations}
\end{figure}

Two kink solutions were launched with at differ initial points
in the junction with intial velocity $v=0.6$. An dc bias of more
than $\gamma=0.1$ is needed to support motion of two fluxons in
the junction. For $\gamma =0.1$ it is observed that only a
single fluxon propagates through the junction as can be seen
from Fig. \ref{annihilations}(a). A dc bias of $0.12-0.45$
supports two fluxon propagation in the junction in the absence
of an ac biasing. However if an ac bias of 0.1 is applied along
with $\gamma=0.41$ creation of a fluxon occurs as shown in Fig.
\ref{annihilations} (b).

\subsection{Third Fiske Step}
\begin{figure}[tbh]
\centering
\includegraphics[width=6cm]{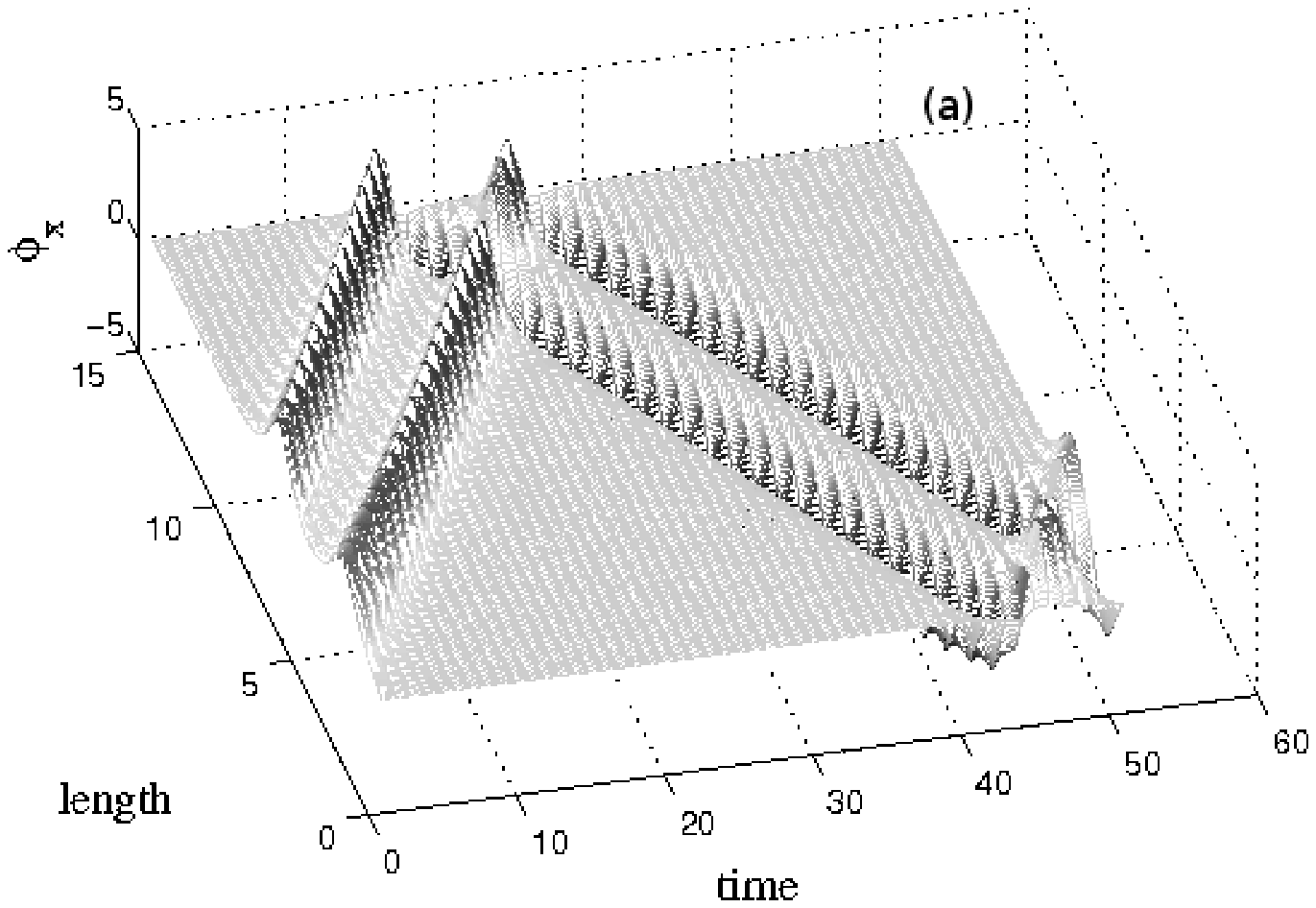}
\hspace{1cm}
\includegraphics[width=6cm]{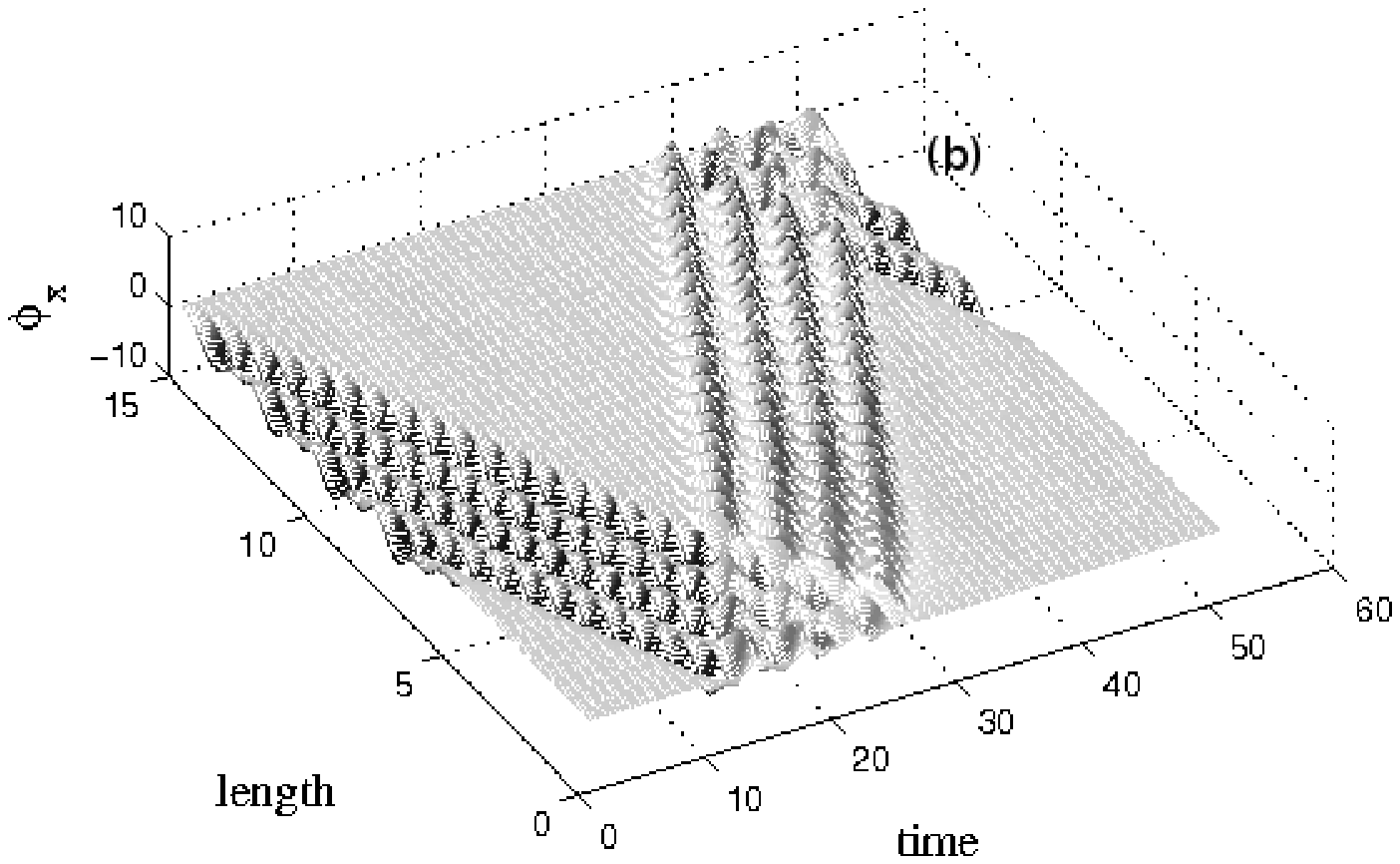}
\caption{The pattern shows two fluxons propagating in a JJ with
l=15.
Other parameter values are $idc=0.3$(b)Creation of fourth fluxon $idc=0.55$} %
\label{annihilationt}
\end{figure}

In this case, three kink fluxons are launched at different
initial points. A dc bias of $\gamma =0.4$ is needed to support
the three fluxon propagation in the junction. In
Fig.\ref{annihilationt}(a)  we numerically show that  only two
fluxons propagate through the junction for a $\gamma$ value of
$0.3$.  Also for  $\gamma = 0.4$, if an ac bias is applied
annihilation of one fluxon occurs again giving the two solitonic
propagation as shown in \ref{annihilationt}(a). The ac biasing
causes annihilation and if the $i_0$ value is increased to
$0.19$ or more the solitonic profile is lost.

Creation of fluxon is observed for $\gamma$ values of 0.55 as
shown in Fig. \ref{annihilationt}(b) with ac biasing destroying
the structure even for $i_0=0.1.$ it is to be noted that all
these effects takes place only in the presence of an external
magnetic field in semiannular JJs. In the absence of magnetic
fields, we were not able to observe creationa nd annihilation of
fluxons.

\section{Conclusions}
We have studied the dynamics of a fluxon trapped in  a semi
annular JJ in the presence of an external magnetic field along
with an ac biasing. This method of applying ac biasing offers a
much easier and controllable way to induce a harmonic periodic
modulation to the junction. In the presence of an external
magnetic field the vortex  remains pinned in the potential well.
The ac biasing modulates the form of the potential and we obtain
an oscillating potential with frequency of oscillation equal to
the driving field. In the presence of an ac-drive and magnetic
field , fluxon creation and annihilation phenomena is observed.
This has been demonstrated for one, two and three fluxons and
can be extended to higher number solutions. The fluxon creation
and annihilation process being crucial for the understanding of
the internal dynamics of the junctions, it will have important
applications in design and fabrication of superconducting
digital devices.
\section*{Acknowledgments}

The authors acknowledge DRDO, Government of India for financial
assistance in the form of a major project.

\end{document}